\documentclass[9pt,twocolumn,twoside]{pnas-new}
\usepackage{ulem}
\templatetype{pnasresearcharticle}

\newcommand{\COMMENTED}[1]{}

\title{Tuning the quantumness of simple Bose systems: A universal phase diagram}

\author[a,1]{Youssef Kora}
\author[a]{Massimo Boninsegni} 
\author[b]{Dam Thanh Son}
\author[c,d]{Shiwei Zhang}

\affil[a]{Department of Physics, University of Alberta, Edmonton, Alberta, T6G 2E1, Canada}
\affil[b]{Kadanoff Center for Theoretical Physics, University of Chicago, Chicago, IL 60637}
\affil[c]{Center for Computational Quantum Physics, Flatiron Institute, New York, New York 10010}
\affil[d]{Department of Physics, The College of William and Mary, Williamsburg, Virginia 23187}

\leadauthor{Kora} 

\significancestatement{Predicting the properties of a quantum-mechanical system of many interacting particles is a major goal of modern science, and an
outstanding challenge. We consider a compact and versatile model which captures the essential features of a broad class of systems made of particles obeying 
Bose-Einstein statistics, and which allows one to systematically dial up the effect of quantum entanglement in the presence of particle interaction by tuning a single parameter.  We are able to obtain exact numerical results for the phase diagrams of these systems. Possible directions for experimental realization of the predictions are discussed.}

\authordeclaration{Authors declare no competing interests.}
\correspondingauthor{\textsuperscript{1}To whom correspondence should be addressed. E-mails: dtson@uchicago.edu, ykora@ualberta.ca}


\begin{abstract}
We present a comprehensive theoretical study of the phase diagram of a system of many Bose particles 
interacting with a two-body central potential of 
the so-called  Lennard-Jones form. 
First-principles path-integral 
computations are carried out, providing essentially 
exact numerical results on the thermodynamic properties. The theoretical model used here provides a realistic and remarkably general framework for describing simple Bose systems 
ranging from crystals to 
normal fluids to superfluids and gases. 
The interplay between particle interactions on the one hand, quantum indistinguishability and delocalization on the other, is characterized by a single 
{\em quantumness} parameter, which can be  tuned
to engineer and
explore different regimes. Taking advantage of the rare combination of the 
versatility of the many-body Hamiltonian and 
the possibility for exact computations, 
 we systematically investigate the phases of the systems
 as a function of pressure ($P$) and temperature ($T$), as well as 
 the quantumness parameter. We show how the topology 
of the phase diagram evolves from the known case of $^4$He, as the system 
is made more (and less) quantum,  and
compare our predictions with available results from mean-field theory.
 Possible realization and observation of the phases and physical regimes predicted here 
 are discussed in various experimental systems, 
 including hypothetical muonic matter. 
\end{abstract}

\dates{This manuscript was compiled on \today}
\doi{\url{www.pnas.org/cgi/doi/10.1073/pnas.XXXXXXXXXX}}

\begin{document}

\maketitle
\thispagestyle{firststyle}
\ifthenelse{\boolean{shortarticle}}{\ifthenelse{\boolean{singlecolumn}}{\abscontentformatted}{\abscontent}}{}

\dropcap{O}ne of the major themes of modern physics is the prediction of macroscopic properties and phases of thermodynamic assemblies of atoms and molecules directly from first principles.
A famous quote by Weisskopf from 1977 captures the aspiration, and also underscores the challenge: ``Assume that a group of
intelligent theoretical physicists have lived in closed buildings from birth that
they never had occasion to see natural structures... What would they be able to
predict from a fundamental knowledge of quantum mechanics? They would predict the
existence of atoms, of molecules, of solid crystals, both metals and insulators,
of gases, but most likely not the existence of liquids.'' \cite{Weisskopf1977}.
Although Weisskopf focused on liquids, his remark 
highlighted the broader difficulty in treating inter-particle 
interactions and emergent phenomena. 
Interestingly, factoring in the possibility of computer simulations would almost certainly have changed this assessment. Simulations using simple models of atomic
interactions 
allow one to make predictions of  equilibrium structure and thermodynamic properties of many simple systems, including those of liquids. 

The rapid increase of modern computing power and development of computational algorithms have greatly expanded the role 
of computer simulations and computation, now encompassing many subareas of physics, chemistry, materials science, etc. 
Despite early fears, expressed, e.g., by Dirac that ``the exact application of these laws leads to equations much too complicated to be soluble,'' \cite{Dirac1929}, we are now in the position to apply the fundamental laws of quantum mechanics to 
 a large number of many-body systems, with precision sufficient for fruitful comparison with experiment.
Of particular interest in the understanding of how quantum-mechanical effects alter the qualitative behavior of the system predicted classically. 
For systems obeying Fermi statistics, 
it is not yet possible to \emph{systematically} reach the accuracy  necessary for reliable 
predictions of new reactions, new structures, or new
phases of matter; indeed, this remains a grand challenge.
However, if the constituent particles  obey Bose statistics  
it is now possible in principle to obtain exact numerical estimates of thermodynamic averages of relevant physical observables, for many relevant physical systems.

A broad class  of condensed matter systems is well characterized by pair-wise, central interactions among constituent particles (e.g., atoms), featuring {\em a}) a  strong repulsion at short inter-particle separations (from Pauli exclusion principle, acting to prevent electrons from different atomic or molecular clouds from overlapping spatially) {\em b}) a weak attractive tail at long distances, arising from mutually induced electric dipole moments. 
A  widely used approximate model to describe such an interaction is the Lennard-Jones (LJ)  
potential:
\begin{equation}
V_{LJ}(r) = 4 \epsilon \left[\left( \frac\sigma r \right)^{12} - \left(\frac\sigma r \right)^6 \right],
\end{equation} 
where $\epsilon$ is the depth of the attractive well, $\sigma$ is the characteristic range of the interaction, and $r$ is the separation between the two particles. 
Despite its simplicity, the LJ potential effectively 
accounts for the physical behavior of a large number of simple liquids.
\\ \indent
A chief example of this type of condensed matter system is helium. 
Helium is unique among all substances, in that it 
does not
solidify at low temperature, under the pressure of its own vapor. Its most common isotope, $^4$He, undergoes a transition to a superfluid phase at a temperature of $2.17$ K.
Both the fact that no crystallization occurs
and the superfluid transition 
are understood as consequences of Bose statistics \cite{feynman,exchanges}, which $^4$He atoms (composite particles  of zero total spin) obey. 
At higher temperature, $^4$He shows a behavior typical of other fluids, e.g., it has a liquid-gas critical point at temperature about 5.19 K and pressure 227 kPa.
\\ \indent
 The question immediately arises of how general some of the properties of $^4$He are among Bose systems featuring the same kind of interaction, or 
how they might evolve with the mass of the particles and the interaction parameters, or whether new phases
might appear.
 \\ \indent
A theoretical description of a system of interacting bosons based on the LJ potential constitutes a simple but remarkably general framework in which 
such questions can be addressed.
On taking $\epsilon$ ($\sigma$) as our unit of energy (length), the Hamiltonian is fully parametrized by
the dimensionless parameter\footnote{The $\Lambda$ parameter used here is proportional to the square of the well-known De Boer parameter \cite{deboer}}
\begin{equation}
\Lambda = \frac{\hbar^2}{ m\epsilon\sigma^2}\,,    
\end{equation}
whose magnitude expresses the relative importance of the kinetic and potential energies.
The larger the value of $\Lambda$, the more significant the 
quantum effects in the dynamics of the particles, and the higher the temperature
to which they can be expected to persist. 
Conversely, in the $\Lambda\to 0$ limit the potential energy dominates, and the behavior of the system is largely classical.
 \\ \indent 
In order to make this argument more quantitative, we note that for $^4$He, $\epsilon\equiv\epsilon_{\rm He}=10.22$\,K and $\sigma\equiv\sigma_{\rm He}=2.556$\,\AA, i.e., $\Lambda=0.18$, which is the second highest value among naturally occurring substances (the highest being 0.24 for the lighter helium isotope, $^3$He, a fermion). For comparison, for a fluid of parahydrogen molecules,  i.e., spin-zero bosons of mass one half of that of a $^4$He atom, $\epsilon=34.16$\,K and $\sigma=2.96$\,\AA, yielding $\Lambda=0.08$. In stark contrast to helium, fluid parahydrogen crystallizes at a temperature $T$=13.8\,K, well above that at which Bose-Einstein condensation might take place. Although quantum effects are observable \cite {dusseault} near melting, there is no evidence of a superfluid phase, even in reduced dimensions, where quantum effects are amplified \cite{1db}.
\\ \indent
One might wonder what the phase diagram may be if $\Lambda$ should be significantly greater than the helium value of $\sim 0.2$.
This may seem like a purely academic question,
given that helium is an ``outlier'' among naturally occurring substances.  
However, there are avenues that may allow
experimental realizations of LJ Bose systems with 
larger $\Lambda$ values.
Confined assemblies of ultracold atoms, in which the interaction can be ``tuned" by means of techniques such as the Feshbach resonance (see, for instance, Ref.~\cite{feshbach}), may provide a test for some of the predictions, at least in the low density limit. 
In excitonic systems in solids, it may be possible 
to engineer the effective mass of holes to affect the effective value of 
$\Lambda$ for the multi-exciton system.
The recent flourish of activities in flat-band materials~\cite{Cao2018-superconductor,Cao2018-insulator} could also result in the ability of fashioning 
effective interactions, i.e., another way to tune $\Lambda$.
\\ \indent

Moreover, there are intriguing possibilities with exotic atoms,  in which one or more electrons are replaced by other subatomic particles of the same charge, such as muons \cite{Egan}; recently, a long-lived ``pionic helium'' has been created \cite{pionic}. A more radical approach consists of replacing {\em all} electrons \cite{tajima,wheeler}; for example,
a ``muonic'' version of a given element of mass $M$ has an equivalent mass $M_{eq}$ given by \cite{son2020phase}
\begin{equation}\label{muonic}
M_{eq} =  \left(1+ \frac{Z}{A}\frac{m_\mu}{m_N}  \right) \frac{m_e}{m_\mu}M
\end{equation}
where $m_\mu$ and $m_e$ are the masses of the muon and the electron respectively.
The replacement of electrons by muons causes {\em a}) a shrinkage of the range ($\sigma$)  of the inter-particle potential by a factor of $m_\mu/m_e$
($\sim 200$) and {\em b}) an increase in the well depth ($\epsilon$) by the same factor, 
resulting in a 200-fold increase of $\Lambda$ ---
sufficient to bring even systems made of heavier elements, e.g., Ne, whose condensed phase displays essentially classical physical behavior, into the highly quantum regime.
\\ \indent
In this work, we perform a comprehensive study of the
universal phase diagram of LJ Bose systems.
We use state-of-the-art quantum Monte Carlo (QMC) methods to compute numerically exact thermodynamic averages of relevant physical observables at finite temperatures.
Given the presence of both strong interactions and large quantum effects in these systems, systematically accurate many-body computations 
are crucial for reliable predictions. 
We map out  the complete thermodynamic phase diagram 
as a function of pressure and temperature, varying
the  parameter $\Lambda$ to explore a variety of physical regimes
ranging from almost entirely classical to the ultra-quantum. 
\\ \indent
The remainder of this paper is organized as follows: in Sec.~\ref{the} we describe the model of the system, and briefly summarize the methodology we utilized. In Sec.~\ref{res}, we present and discuss our results
in several subsections separated by the different 
regimes of $\Lambda$, and we finally outline our conclusions in Sec.~\ref{conc}.
\\ \indent

\section{Theoretical framework}\label{the}

\subsection{Model}\label{mo}
We consider an ensemble of $N$ identical particles of mass $m$ obeying Bose statistics, enclosed in a cubic box of volume $V$, with periodic boundary conditions in the three directions. The density of the system is therefore $\rho=N/V$. Particles interact via the LJ potential. 
As mentioned in the introduction, we take the characteristic length 
$\sigma$ as our unit of length, and the well depth $\epsilon$ as that of energy.
The
dimensionless quantum-mechanical many-body Hamiltonian reads as follows:
\begin{equation}\label{hamiltonian}
\hat H = - \frac{1}{2} \Lambda \sum_{i}^N\nabla^2_{i} + 4\sum_{i < j}^N 
 \left(\frac{1}{r_{ij}^{12}} - \frac{1}{r_{ij}^{6}}\right)\,,
\end{equation}
where the first (second) sum runs over all particles (pairs of particles), and 
$r_{ij}\equiv |{\mathbf r}_i-{\mathbf r}_j|$ is the distance between particles $i$ and $j$. 
In these reduced units, $\Lambda$ is the only parameter of the Hamiltonian, and therefore its numerical value univocally determines the nature of its equilibrium phase, at any values of pressure and temperature.
\\ \indent
In discussing our results, we shall at times find it useful to refer to a particular system not in terms of its value of $\Lambda$, but rather 
of its ``equivalent helium mass'' $X$, defined as the mass of a hypothetical helium isotope ($^X$He, always assumed to be a boson)
which yields the same value of $\Lambda$.
That is, a system of mass $M$ characterized by 
LJ interaction parameters 
of $\epsilon$ and $\sigma$ has
\begin{equation}\label{meq}
X = M \frac{\epsilon}{\epsilon_{\rm He}} \left(\frac{\sigma}{\sigma_{\rm He}} \right)^2\,.
\end{equation}
Thus, the mathematical description of the system can also be equivalently parametrized in terms of $X$, instead of $\Lambda$. \\ 

\subsection{Methodology}\label{me}

\begin{figure}[h]
\centering
\includegraphics[width=1\linewidth]{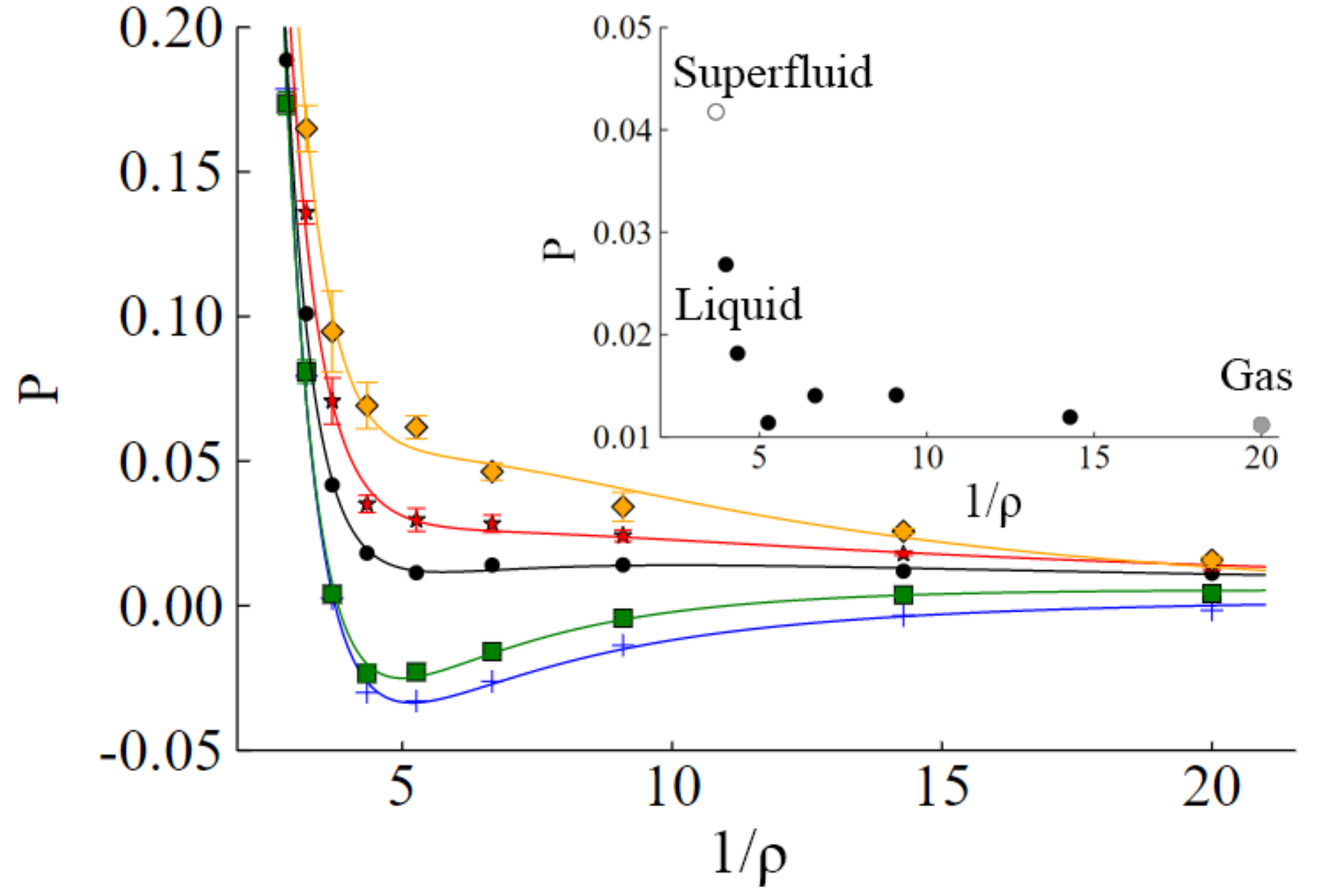}
\caption{The pressure as a function of specific volume at temperatures of 0.1 (crosses), 0.2 (squares), 0.32 (circles), 0.4 (stars), 0.5 (diamonds). This serves as a tool for detecting coexistence between two phases of different densities, as explained in the text. This particular result is for LJ boson $^3$He. Inset: Same as main graph but with a smaller $P$ scale and only $T=0.32$. Different symbols distinguish the superfluid phase (empty), the normal phase (filled), and the gas phase (grey).}
\label{he3iso}
\end{figure}

As mentioned above, we 
carry out systematic many-body calculations
of the system described in subsection \ref{mo} using QMC simulations. Specifically, we make use of the well-established continuous-space worm algorithm \cite{worm1,worm2}. The technical details of this methodology are extensively illustrated in the literature, and hence we will not be repeating them here, referring instead the reader to the original references. We utilized a canonical variant of the algorithm in which the total number of particles $N$ is held constant, in order to simulate the system at fixed density \cite{mezz1,mezz2}.
\\ \indent
Note that, while there has been considerable simulation work on classical LJ fluids, previous work on quantum systems
has been mostly limited to  variational ground state studies
\cite{nosanow75,nosanow77}. Indeed, the pioneering simulations of the superfluid transition in $^4$He were based on a more accurate interatomic pair potential \cite{pollock, ceperley95}. Finite temperature QMC simulations of LJ systems have also been performed of solids, in which quantum statistics is neglected \cite{cuccoli,nielaba} on account of the relative infrequency of quantum  exchanges. 
\\ \indent
Although our technique is 
based on the finite-temperature path-integral technique \cite{feynmanhibbs}, the ground state physics is explored by reaching 
sufficiently low $T$ so that the results can be regarded as essentially for
the ground state. Once the low-temperature limit is reached, the equation of state of the system is calculated by computing the energy as a function of density, and the minimum of this function is taken to be the equilibrium density, i.e., the density at which the self-bound system exists at $T=0$. As mentioned in subsection \ref{mo}, the finite temperature physics is readily accessible upon raising the temperature $T$, and the over(under)-pressurized system is explored by raising (lowering) the density $\rho$. The value of the pressure at any given 
$T$ and $\rho$ is calculated through the virial theorem (see, for instance, Ref.~\cite {imada}). In this fashion, one may survey the pressure-temperature phase diagram of the system and explore the different phases thereof.
\\ \indent
We performed simulations for values of the equivalent helium mass $1 \le X \le 8$. Details of the simulation are standard; we made use of the fourth-order approximation for the high-temperature density matrix (see, for instance, Ref. \cite{boninsegni05}), and all of the results quoted here are extrapolated to the limit of time step $\tau\to 0$. 
\\ \indent
 Superfluid order is detected through the direct calculation of the superfluid fraction through the well-established winding number estimator \cite{pollock}. The  superfluid transition temperature is estimated by performing finite size scaling analysis of the results for the superfluid fraction, which requires simulating systems of significantly different sizes \cite{worm2}. We obtained estimates for systems comprising a number of particles  ranging from $N=32$ to $N=512$. Crystalline order in the system is detected through i) visual inspection of the imaginary-time paths and ii) the calculation of the pair-correlation function. For computational convenience, we simulated all crystalline phases assuming a body-centered cubic structure. It is known that the energy difference between that and the hexagonal close-packed, in which, e.g., $^4$He crystallizes under pressure, is small (less than 0.02 $\epsilon$ in $^3$He \cite{mpfb}).
\\ \indent
The liquid-gas critical temperature is inferred indirectly, through the computation of the pressure as a function of volume at different temperatures. By definition, the critical temperature is the highest temperature for which there is coexistence between the liquid and gas phases, which has the signature of a flat region in the pressure-volume isotherm. This behavior, however, only occurs in an infinite system. In the case of finite systems accessible to numerical simulations, in which separation into two coexisting phases is energetically  unfavorable, this behavior is reflected by the system acquiring negative compressibility, i.e., the isotherm showing 
positive slope in the coexistence region \cite{pathria}. By plotting isotherms at different temperatures, one can identify the liquid-gas critical temperature as the highest temperature for which there is evidence of coexistence. This is illustrated shown in Fig.~\ref{he3iso} with an example. 
\\ \indent
\begin{figure}[h]
\centering
\includegraphics[width=1\linewidth]{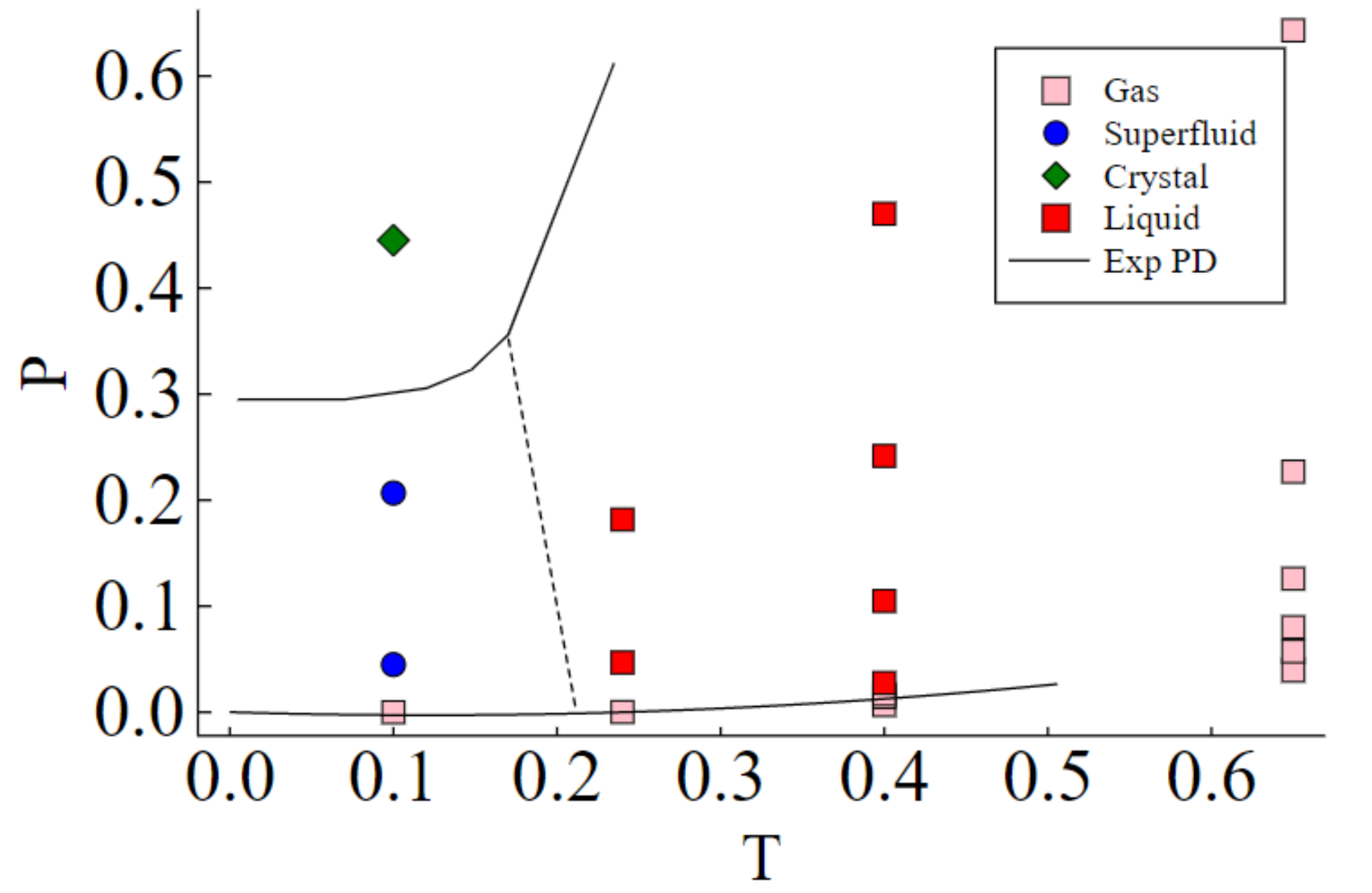}
\caption{The pressure-temperature phase diagram of LJ $^4$He. Solid and dashed lines represent the experimentally determined phase boundaries. Solid lines correspond to first order transition, and dashed to second order.}
\label{he4}
\end{figure}
As a first gauge of the accuracy and reliability of our approach, we  
study $^4$He (i.e., $\Lambda=$ 0.1815).
The topology of the $P$-$T$ phase diagram of $^4$He
is well-known from a wealth of experimental measurements \cite{wilks,barenghi} and theoretical studies \cite{ceperley95}
throughout the decades, with which we can compare our results.
Although most microscopic calculations \cite{ceperley95} of helium utilize the more accurate Aziz pair potential \cite{aziz}, the LJ potential is known to give an
excellent approximation in $^4$He.
Additionally, three-body terms have been shown \cite{mpfb} to account for a relatively small correction to the thermodynamic equation of state, 
with insignificant effect on structural or superfluid properties. 
Comparing our results for $X$=4 against experimental 
phase boundaries, as shown in Fig.~\ref{he4},
thus serves as validation for our methodology. 
\\ \indent
This phase diagram 
features two critical temperatures: i) the superfluid transition temperature $T_\lambda$, and ii) and the temperature that marks the end of the liquid-gas coexistence line $T_{\rm LG}$, i.e., the highest temperature at which there is a phase transition between a liquid phase and a gas phase. Clearly, $T_{\rm LG} > T_{\lambda}$ in this case. However, as one continues to lower the value of the mass, one expects i) quantum-mechanical effects to become more prominent, thus enhancing superfluidity and raising the value of $ T_{\lambda}$, and ii) zero-point motion to increasingly dominate the potential energy, causing the system to become less bound and suppressing the liquid phase, causing $T_{\rm LG}$ to go down. 

We systematically investigate these trends in Sec.~\ref{res}.

\section{Results}\label{res}

\subsection{Overview}\label{overview}

\begin{figure}[h]
\centering
\includegraphics[width=1\linewidth]{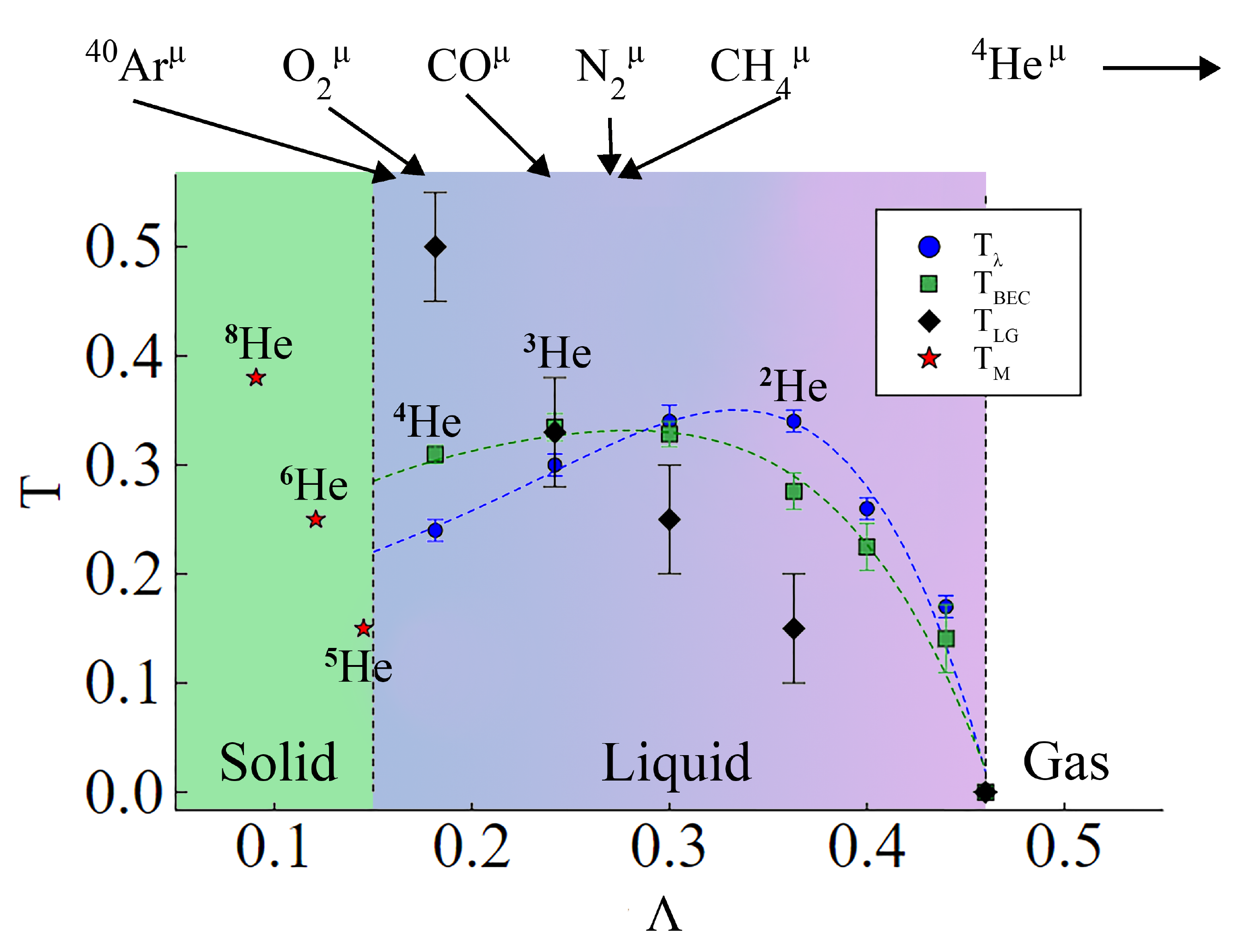}
\caption{Ground state and liquid-gas critical temperature of the system as a function of $\Lambda$. The liquid-gas critical temperature ($T_{\rm LG}$, diamonds) is  determined by the procedure discussed 
in Sec.~1.B and illustrated for $X=3$ in Fig.~\ref{he3iso}. Also shown are the superfluid ($T_{\rm \lambda}$, circles) and Bose-Einstein ($T_{\rm BEC}$, boxes) transition temperatures of homogeneous fluids, as well as the melting temperatures ($T_{\rm M}$, stars) of crystals. $T_{\rm \lambda}$, $T_{\rm BEC}$ and $T_{\rm M}$ are computed by holding the density fixed at the ground state equilibrium value. When not shown, statistical uncertainties are smaller than the size of the symbols. Lines are guides to the eye.}

\label{mainpd}
\end{figure}

\begin{figure}[h]
\centering
\includegraphics[width=1\linewidth]{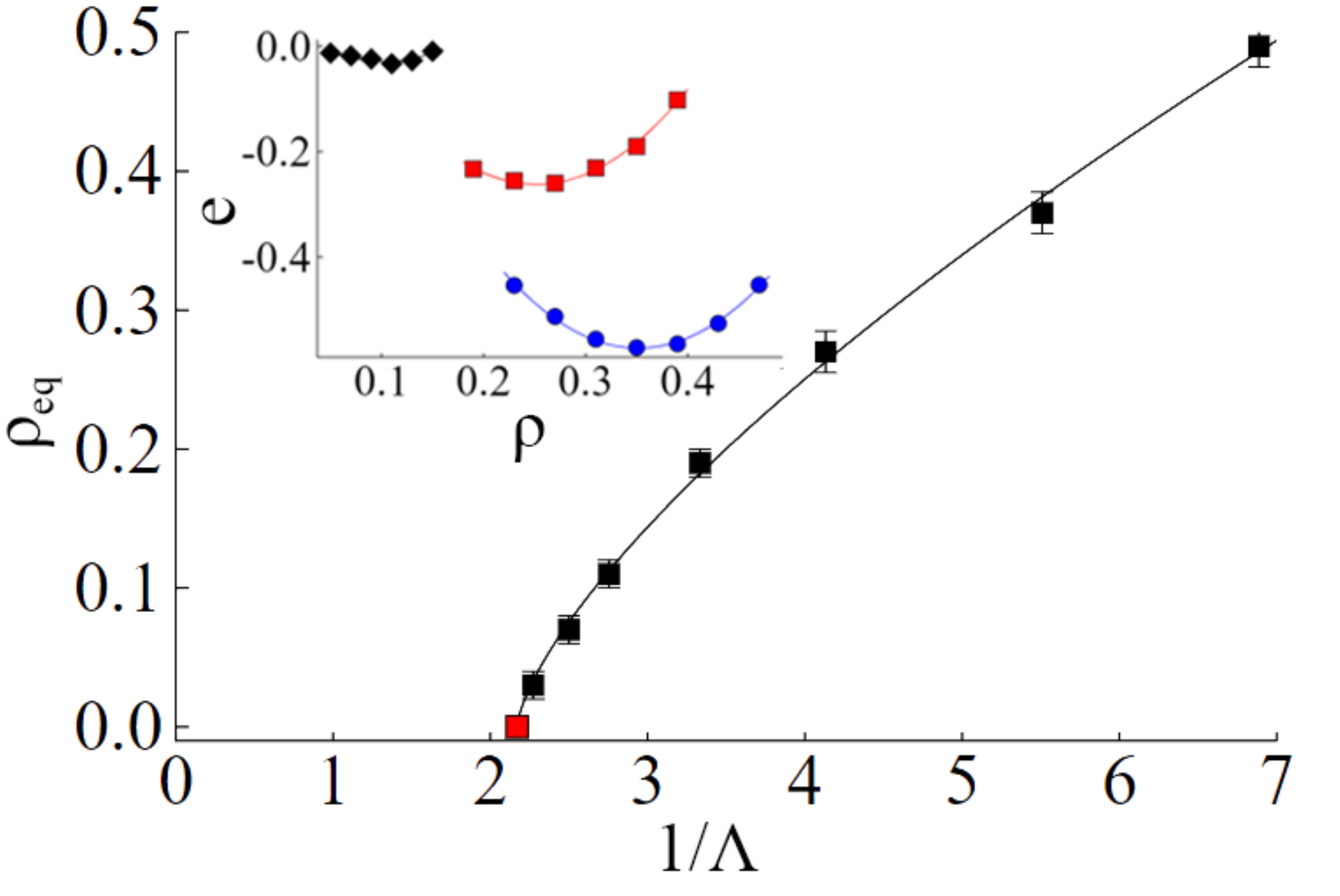}
\caption{The ground state equilibrium density of the system as a function of the inverse de Boer parameter. 
The intercept at $\rho_{\rm eq}=0$ shows 
the minimum nuclear mass that remains self-bound at zero temperature;
the red square is an exact result obtained from 
the two-body scattering length \cite{gomez,zwerger}. 
Inset: examples of the equation of state at $\Lambda=0.1815,0.242,0.363$, respectively from bottom to top. 
}
\label{eqdens}
\end{figure}

The results of our extensive QMC computations 
are summarized in Fig. \ref{mainpd}. In this subsection, we
discuss the main ground state features of this diagram and give a brief overview of the different physical regimes at zero temperature, before moving on to describe finite temperature characteristics. 
The various transition temperatures in Fig.~\ref{mainpd} 
were computed at specific values of $\Lambda$. The corresponding 
equivalent helium mass $X$ values are also shown in 
the figure.
We indicate with arrows the locations the muonic counterparts of some molecules. \\ \indent
The different shades in Fig.~\ref{mainpd} represent the different ground states of the system, depending on the value of $\Lambda$. Three distinct physical regimes can be identified. At low values of $\Lambda$ (high values of the nuclear mass), the ground state is a crystal. 
At a value of $\Lambda \approx 0.15$, the system
quantum melts into a superfluid that remains self-bound.
As $\Lambda$ is further increased, the binding is weakened. 
This behavior is illustrated in Fig.~\ref{eqdens}, which shows the equilibrium density going down as $\Lambda$ grows, to finally hit zero upon reaching 
another critical value $\Lambda_c$,
whereupon the system undergoes quantum unbinding.
In the regime $\Lambda > \Lambda_c$, the ground state is a superfluid gas. 
\\ \indent
From the many-body equation-of-state results, we obtain an estimate of
$\Lambda_c \approx 0.46$ as shown in Fig.~\ref{eqdens}, which corresponds to $X \approx 1.6$.
This result agrees
with an earlier prediction made in Ref.~\cite{zwerger}
using the zeros of the two-body scattering length \cite{gomez}, 
confirming the argument based on few-body considerations.
In the series expansion of the effective potential in terms of a classical field, the three-body term has the opposite sign with respect to the two-body term, as we approach  $\Lambda_c$ from below. 
We can also compute the three-body scattering hypervolume $D$, related to the three-body coefficient $\lambda_3 /3$ by $\hbar^2 D/6m = \lambda_3/3$.
Our estimate is obtained by fitting the energy as a function of density at 
a value (chosen to be $\Lambda=0.44$)
close to $\Lambda_c$ with a third degree polynomial and extracting the value of the coefficient of the cubic term. This gives $D/\sigma^4 = 57 \pm 8$, which is again consistent with the estimates made from few-body calculations in Refs.~\cite{zwerger,mestrom}.
\\ \indent
The finite-temperature behavior of all three physical regimes is also shown in Fig.~\ref{mainpd}. The crystalline phase 
melts into a non-superfluid liquid
upon increase of 
the temperature.
This is not surprising,
and underscores the importance of quantum-mechanical exchanges, which underlie superfluidity, in the melting of the Bose solid \cite{exchanges}. Melting occurs at a temperature which decreases on increasing the value of $\Lambda$. In the Liquid, we computed three different temperatures: i) the liquid-gas critical temperature, ii) the superfluid transition temperature at the ground state equilibrium density, iii) the Bose-Einstein condensation temperature of the non-interacting system $T_{\rm BEC}\approx 3.3125\ \Lambda\ \rho^{2/3}$, also at the ground-state equilibrium density. The interplay between the three temperatures is plotted in Fig. \ref{mainpd}, and is discussed in more detail in subsection \ref{intmass}. Finally, we have the superfluid gas regime, in which the system behaves very similarly to a
dilute Bose gas. 
\\ \indent
In the following three sub-sections, we provide more detailed descriptions of the different physical regimes in Fig.~\ref{mainpd}, moving from smaller to larger values
of $\Lambda$. Detailed
$P$-$T$ phase diagrams are computed at representative $\Lambda$ values to probe the different phases and 
the topology of the phase transitions.
It is important to reiterate that our results are all using
the simple LJ atom-atom interaction. Despite its generality, there will be 
situations, for example low-density diatomic gases or very high pressure states,
where new phases emerge which are not captured by our Hamiltonian.

\subsection{Low $\Lambda$ regime}\label{highmass}

\begin{figure}[h]
\centering
\includegraphics[width=1\linewidth]{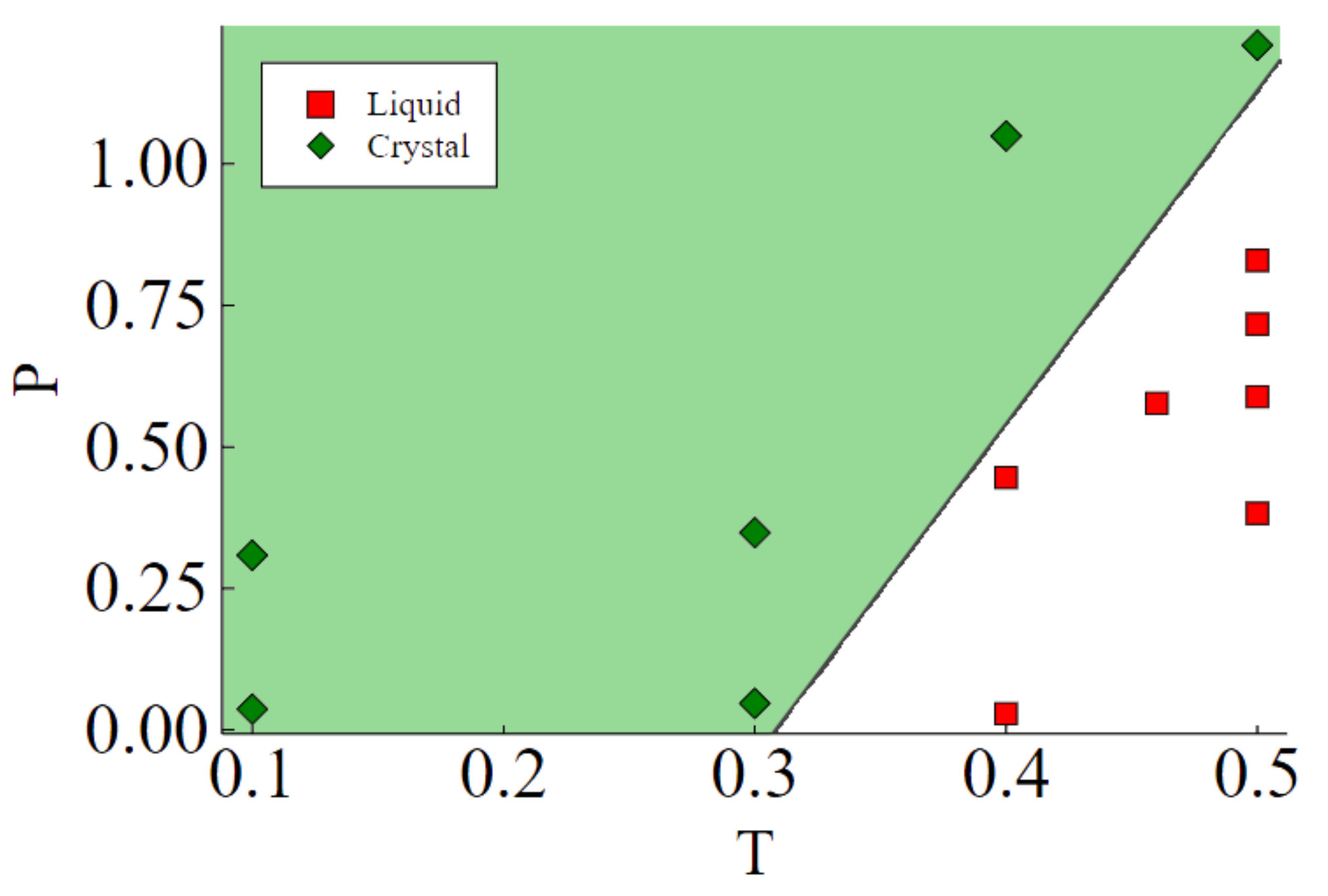}
\caption{The pressure-temperature phase diagram of LJ boson $^8$He ($\Lambda=0.09075$). The solid line, drawn as a guide to the eye, corresponds to the first order transition.
The low-pressure vapor region is not visible on this scale.
 }
\label{he8}
\end{figure}

At values of $\Lambda$ corresponding to $X>4.8$,
quantum mechanical exchanges  are suppressed, and
the ground state is primarily the result of minimizing the potential energy, i.e., a crystal. Nevertheless, as shown in Ref.~\cite{exchanges}, exchanges play a crucial role in the determination of the melting temperature.
(It is worth noting that the isotopes $^8$He and $^6$He have both been realized in the laboratory,
with single nuclei half-lives of 0.12\,s and  0.8\,s, respectively.)

The pressure-temperature phase diagram at $\Lambda=0.09075$, corresponding to $X=8$, is shown in Fig. \ref{he8}.
\\ \indent
The melting temperature of the equilibrium crystalline phase goes down as $\Lambda$ grows, as shown in Fig.~\ref{mainpd}. In particular, we estimate the melting temperature of $^6$He to be about $2.5$ K, the lowest among naturally occurring substances.  
\\ \indent
On  raising $\Lambda$, one encounters the fictitious boson isotope $^5$He, which still lies on the solid side of the solid-liquid boundary. Here, the crystalline ground state remains stable against quantum fluctuations, albeit with a relatively small melting temperature of $\sim 1.5$ K. Interestingly, for a value of $X$ this close to the solid-liquid boundary, we find a possibly long-lived, over-pressurized superfluid phase at the equilibrium density. Such a phase is not realized in systems deeper within the classical regime, such as parahydrogen \cite{boninsegni18}, as well as $^6$He and $^8$He. It is reminiscent of the situation of $^4$He, in which it is possible to achieve metastable superfluid phases at pressures much higher than the crystallization pressure \cite{balibar,wern,condensate,superglass}.

\subsection{Intermediate $\Lambda$ regime}\label{intmass}
As the value of $\Lambda$ further grows, one crosses the solid-liquid boundary and encounters the well-characterized $^4$He, the results for which, as mentioned in section \ref{me}, serve as validation for our methodology, and are compared against the experimental results in Fig.~\ref{he4}.

\begin{figure*}[t]
\centering
\includegraphics[width=1\linewidth]{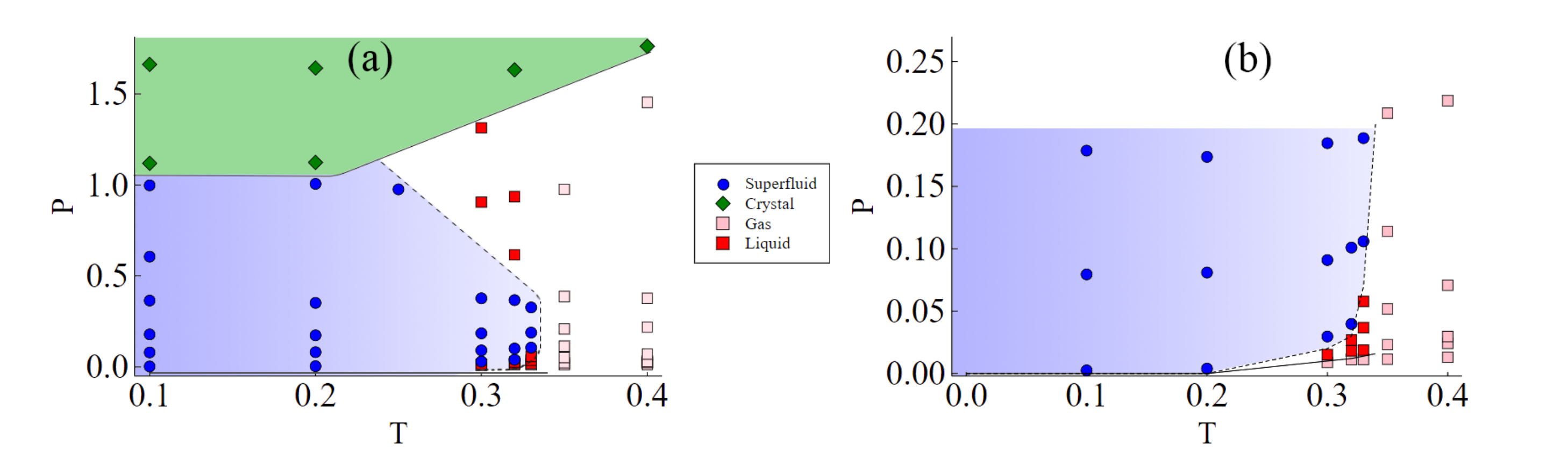}
\caption{(a) The pressure-temperature phase diagram of LJ boson $^3$He ($\Lambda=0.242$). (b) Same as (a) but with a zoom into the lower pressure portion.
Lines are drawn schematically based on the discrete 
data points to guide the eye.
Solid lines correspond to first order transition, and dashed to second order.
 }
\label{he3}
\end{figure*}

As shown in Fig.~\ref{mainpd}, as one moves further to the right, the superfluid transition temperature ($T_{\lambda}$, computed at the ground-state equilibrium density) goes up, then plateaus and goes slightly back down. This behavior is the the result of a competition between two effects that take place as $\Lambda$ grows: i) the system becomes increasingly quantum mechanical, allowing superfluidity to be possible
at higher temperatures, and ii) the equilibrium density decreases,
which means that the particles are on average more widely spaced apart, requiring larger de Broglie wavelengths, and hence lower temperatures, to achieve Bose condensation.
On the other hand, the liquid-gas critical temperature $T_{\rm LG}$ goes down monotonically as the value of $\Lambda$ is increased, as the system experiences more zero-point fluctuations and becomes more loosely bound. 
\\ \indent
$^3$He is located at a point fairly close to the crossing of the two temperatures. 
In Fig.~\ref{he3}, we map out the pressure-temperature phase diagram of 
the fictitious bosonic $^3$He,
which is
distinct from that of $^4$He in a number of different ways.
Overall, the superfluid region expands greatly, 
pushing the superfluid-crystal transition line up to higher $P$, while pushing
the superfluid transition line to the right. 
The first-order line 
that separates the liquid and gases phases shrinks 
significantly as $T_{\rm LG}$ drops from 0.5 to around 0.34.
The second-order line 
that separates the superfluid and normal phases,
aside from moving to higher temperatures as mentioned, behaves quite differently from the monotonic line with negative slope that appears in $^4$He. Instead, the line starts with a very small positive slope at low pressure, bulges out, then curves back around and acquires a negative slope as it approaches the crystalline regime. 
\\ \indent
It is useful to examine more closely the topology of the phase diagram in the vicinity of $T_{\rm LG}$, as shown 
in Fig.~\ref{he3}b. When $\Lambda$ is increased, the second-order line expands to the right 
and its lower part bends toward the first-order line, which shrinks
as its end, the critical point $T_{\rm LG}$, moves to the left. 
Ref.~\cite{son2020phase}
studied the evolution of the phase diagram of the LJ 
Bose liquid in the ultra-quantum regime through mean field considerations based on Landau theory.
The authors argued that a
portion of the second-order line should turn first-order
before the critical point $T_{\rm LG}$ 
can merge onto it,
 in order to prevent the superfluid and liquid-gas order parameters from becoming critical at the same point \cite{son2020phase}.
 Our results show that, if such a scenario occurs, it
 is confined to a tiny portion of the superfluid transition line for very specific values of $\Lambda$, 
 which is challenging to target numerically.
\\ \indent
The $(P,T)$ phase diagram in Fig.~\ref{he3} reveals 
an interesting range of temperatures near $0.32$. At such a temperature, if one starts at zero pressure and continues pressurizing, keeping the temperatures constant, one first encounters a gaseous phase, followed by a normal liquid phase, followed by a superfluid liquid phase, followed by another normal fluid phase, finally followed by a crystalline phase at the highest pressures.
\\ \indent
Another interesting result is the the minimum pressure at which bosonic $^3$He is found to crystallize. As shown in Fig. \ref{he3}b, the minimum crystallization pressure is around 1 in our units, which corresponds to roughly 84 bars in SI units. This is much higher than the minimum pressure at which the real system, obeying Fermi statistics, is experimentally known to crystallize, which is around 30 bars \cite{mills}. This result is of considerable importance, as it shows that quantum statistics indeed play a large role in determining the crystallization pressure of the system \cite{exchanges}. The Fermi system is a non-superfluid liquid at these temperatures, which renders it significantly more susceptible to crystallization. The Bose system, being in the more robust superfluid phase, continues to resist crystallization for much higher pressures. This result is consistent with the prediction made by the variational theory in Ref.~\cite{nosanow75}, in which the authors contend that the solidification pressure of a bosonic $^3$He is greater than that of the Fermi system by at least a factor of 2.

\subsection{High $\Lambda$ regime}\label{lowmass}

\begin{figure}[h]
\centering
\includegraphics[width=1\linewidth]{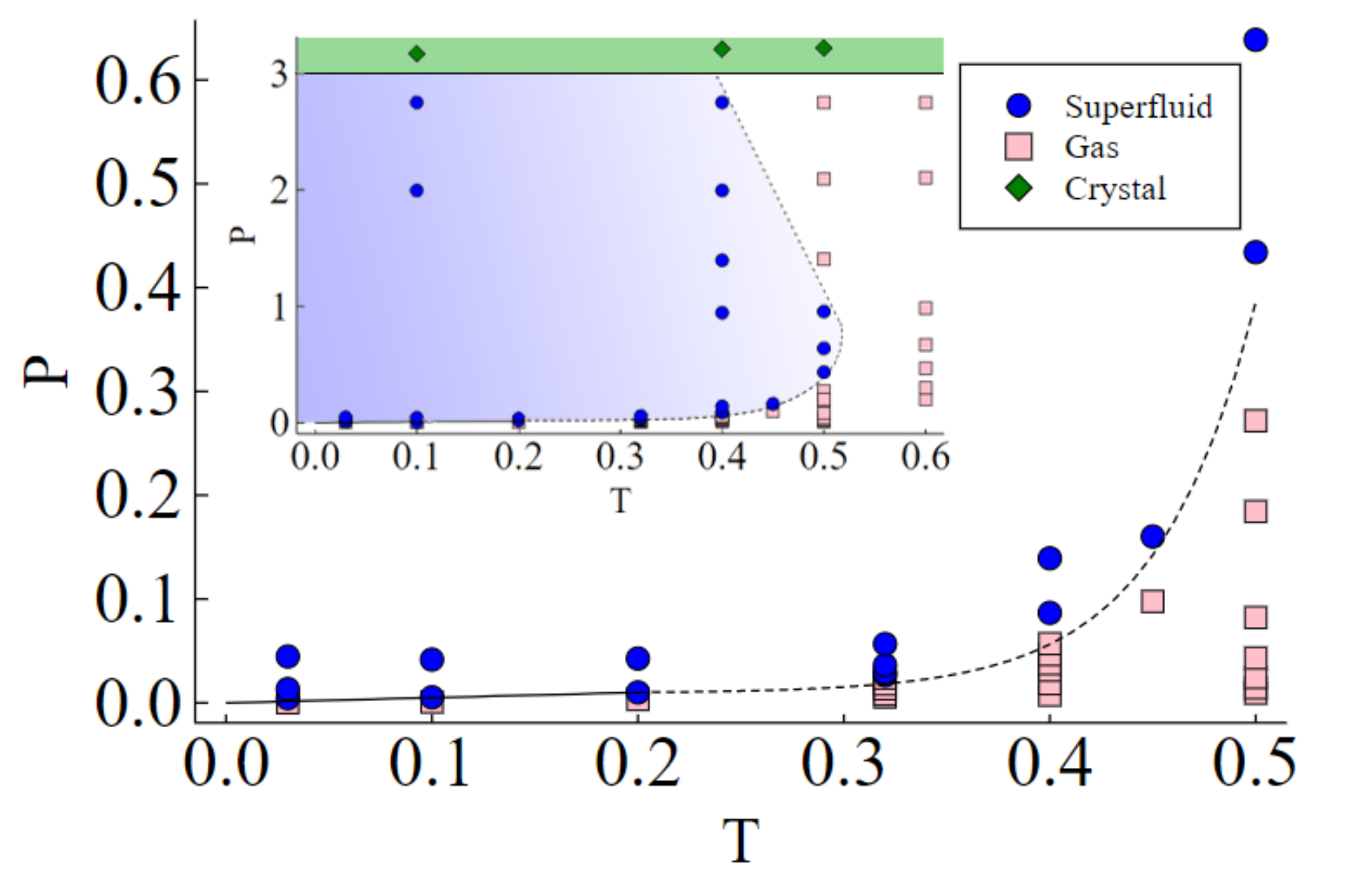}
\caption{The pressure-temperature phase diagram of $^2$He
($\Lambda=0.363$). The main graph shows a zoom of the low pressure
region, while the inset gives a more global view.
Lines are to guide the eye. 
Solid lines correspond to first order phase transition, and dashed to second order. }
\label{he2}
\end{figure}

On further increasing $\Lambda$, one counters $^2$He, which is located at the region where $T_{\lambda}$ exceeds $T_{\rm LG}$. While the system remains self-bound at zero temperature, it boils before losing its superfluidity upon increasing temperature. This is in contrast with
the case in $^4$He which, as the temperature is raised, loses superfluidity long before it boils.
\\ \indent
The pressure-temperature phase diagram for $^2$He is shown in Fig. \ref{he2}, which is simpler 
compared to that of $^3$He. At low temperatures and pressures, a first-order boundary 
separates the superfluid phase and the gas phase. Beyond $T_{\rm LG}$, the phases are separated instead by a second order line, 
which continues to grow as a function of pressure. 
In the inset of Fig.~\ref{he2},
 we present a more complete diagram that includes higher pressures. The behavior at high pressure is similar to that of $^3$He, where the second-order line doubles back and intersects the solid-liquid boundary with a negative slope. 
\\ \indent
As one continues raising the value of $\Lambda$, the first order line separating superfluid and normal phases
progressively  recedes towards the origin, until the system 
no longer features a first-order phase transition. The
 first-order portion vanishes precisely when  $\Lambda=\Lambda_c$, where quantum
unbinding takes place in the ground state, as we
discussed with Fig.~\ref{eqdens}.
For  $\Lambda>\Lambda_c$ there is only a second-order
line separating the superfluid phase and normal gas phase.
These features of the phase boundary 
between the superfluid and gas phases 
at low pressure are correctly captured by 
mean-field and analytic theory  \cite{son2020phase}.

\section{Discussion and Conclusions}\label{conc}

We performed extensive, 
numerically exact many-body computations 
of simple Bose systems interacting through the Lennard-Jones potential, and investigated their physical properties throughout a wide range of the ``quantumness'' parameter $\Lambda$. As a function of $\Lambda$, we studied the evolution of 
the phase diagram, and provided detailed predictions
at several values of $\Lambda$ representative of the 
different physical regimes.
\\ \indent
One goal of our study was
to establish the kind of phases, and phase diagram topology that one can encounter in this very broad class of
systems. Only insulating crystal and (super)fluid phases are present;
no ``supersolid'' is observed, consistent with 
a wealth of theoretical predictions pointing to the absence of a supersolid phase in a system in which the dominant interaction is pair-wise and spherically symmetric and features a ``hard core'' repulsion at short distances \cite{supersolid1,supersolid2}.
No coexistence of two superfluid phases is observed either, which is also consistent with the thermodynamics of the liquid-gas transition and our current understanding of the relation between superfluidity and Bose-Einstein condensation 
in gases.
\\ \indent
Given the generality of the LJ interaction, mapping  out in detail the thermodynamic phase diagram can guide in the design and interpretation of experiments aimed at observing additional phases of matter, as more experimental avenues continue to open up.
Experimental realization of the systems studied here are certainly not limited to helium. 
Among all naturally occurring substances, 
significant quantum effects are observed in parahydrogen, and  can also be expected in two unstable isotopes of helium which possess an even number of nucleons (i.e., they are bosons). 
Higher values of $\Lambda$ may be achieved in a laboratory setting by preparing systems of ultracold atoms,
via exotic matter, 
or in excitonic systems.
\\ \indent
In addition to providing a universal phase diagram to this class of simple Bose system, we hope that 
our investigation also serves as an example of 
the progress to make definitive and comprehensive 
predictions
on interacting quantum many-body systems.
Such examples are still uncommon, but are certainly 
becoming increasingly 
possible, owing to the development
of reliable and robust computational methods and more cross-fertilization 
between them with analytical approaches.

\section*{Acknowledgments} This work was supported by the Natural Sciences and Engineering Research Council of Canada,
a Simons Investigator grant (DTS) and the Simons Collaboration on Ultra-Quantum Matter, which is a grant from the Simons Foundation (651440, DTS).
Computing support of Compute Canada
and of the Flatiron Institute 
are
gratefully acknowledged. 
The Flatiron Institute is a division of the Simons Foundation.

\bibliography{mainbib}

\end{document}